\documentstyle[12pt,aaspp4]{article}

\input{epsf}
\newcommand{\lapprox}{\raisebox{-.5ex}{ $\stackrel{<}{\sim}$\/}}

\begin{document}
\slugcomment{Accepted for ApJ Letters}
\title{The Calibration of the HST Kuiper Belt Object Search:
Setting the Record Straight}
\author{Anita L. Cochran\altaffilmark{1}}
\affil{Astronomy Department \& McDonald Observatory,
University of Texas, Austin, TX~~~78712}
\author{Harold F. Levison\altaffilmark{1}, Peter Tamblyn\altaffilmark{1}, S. Alan Stern\altaffilmark{1}}
\affil{Space Science Department,
Southwest Research Institute, Boulder, CO~~80302}
\author{Martin J. Duncan\altaffilmark{1}}
\affil{Department of Physics, Queen's University,
Kingston, Ontario, Canada K7L 3N6}

\altaffiltext{1}{Based on observations with NASA/ESA Hubble Space 
Telescope obtained at the Space Telescope Science Institute, which is
operated by Universities for Research in Astronomy, Incorporated, under
NASA contract NAS5-26555.}

\begin{abstract}
The limiting magnitude of the HST data set used by Cochran {\it et
al.}~(1995) to detect small objects in the Kuiper belt is reevaluated,
and the methods used are described in detail.  It is shown, by
implanting artificial objects in the original HST images, and
re-reducing the images using our original algorithm, that the limiting
magnitude of our images (as defined by the 50\% detectability limit)
is $V=28.4$.  This value is statistically the same as the value found
in the original analysis.  We find that $\sim50\%$ of the moving
Kuiper belt objects with $V=27.9$ are detected when trailing losses
are included.  In the same data in which these faint objects are
detected, we find that the number of false detections brighter than
$V=28.8$ is less than one per WFPC2 image.  We show that, primarily due
to a zero-point calibration error, but partly due to inadequacies in
modeling the HST'S data noise characteristics and Cochran et al.'s
reduction techniques, Brown et al$.$~1997 underestimate the SNR of
objects in the HST dataset by over a factor of 2, and their
conclusions are therefore invalid.

\end{abstract}


\section{Introduction}

In Cochran {\it et al.}~(1995, hereafter CLSD), we reported the
statistical detection, based on observations obtained with the Hubble
Space Telescope, of a population of small objects which were resident
in the Kuiper belt.  The objects detected had visual magnitude between
27.8 and 28.6, implying radii between $\sim 5$ and $\sim 10\, km$
(assuming an albedo of 4\%).

At the time of publication, the results of CLSD were criticized
on two grounds. The first is that the detections were statistical in
nature; we are unable to fit orbits to our objects.
Indeed, there is a small possibility that our results were a
statistical fluke, a possibility which new, deeper
HST observations will explore.  The second is that the
number of detections did not agree with extrapolations of the
size distribution of large Kuiper belt objects determined from
early ground-based observations (although Weissman \& Levison~1997
showed that the CLSD results were in agreement with the numbers of
Kuiper belt objects needed to populate the known Jupiter-family
comets).  Newer ground-based surveys have changed our
understanding of the size distributions of large objects so
that CLSD results are now consistent with the ground-based
observations by Jewitt {\it et al.}~(1997).

Recently however, Brown {\it et al.} (1997, hereafter BKL) have
contended that the detections reported in CLSD were not possible,
based on an analysis of the noise properties of the data.  BKL
created a simple model of the HST data reduction of CLSD.  They
contended that CLSD could not have detected the objects claimed
because the objects would be overwhelmed by a very large number of
measurements that lie above the detection threshold (hereafter known
as `false detections') due to noise fluctuations.  

In this paper we refute the analysis of BKL by showing that
it is indeed possible to detect objects of $V \sim 28.6$ in the
original CLSD data set without being swamped by
false detections.  Our argument is simple and persuasive.  We
implanted artificial objects in the original CLSD images with
magnitudes between $V=26.5$ and 29.5.  We then reduced these images
using exactly the same procedures used in CLSD, including the same
automated search algorithm to find the objects.  The original four
members of the HST Kuiper belt team performed this analysis for the
original CLSD paper.  As a check of the previous work, we asked
Peter Tamblyn, who has therefore become an author of this paper, to
independently reproduce this analysis.  In his reanalysis, he used the
same algorithm as CLSD, but all his codes and IRAF scripts were
independently developed.  As we show below, this independent
reanalysis agrees with CLSD's original results.

\section{The Test}

  It is well established that one of the most accurate methods for
determining the limiting magnitude of a set of images is to place
artificial objects, of known magnitudes, in the data and search for
them (Mateo~1988; Harris~1990).  This type of procedure has been
routinely used by most observers searching for objects in the Kuiper
Belt (e.g. Luu \& Jewitt~1988; Hainaut et al$.$~1994; Jewitt et
al$.$~1996) and was used by us in CLSD in order to determine the
limiting magnitude of our survey.

  The CLSD data set consists of 34 WFPC2 exposures of a field on the
ecliptic near morning quadrature with the ``Wide V'' (F606W) filter on
HST/WFPC2.  Each exposure was 500--$600\,$sec, for a total integration
time of $5\,$hours.  Thirty hours elapsed from the start of the first
exposure to the end of the last.  For an outer solar system body
observed at quadrature, the Earth's parallactic motion is near zero so
that any apparent motion of the body is attributable to the body's
orbit.  For a body in the Kuiper belt the orbital motion is $\lapprox
1\,$arcsec hr$^{-1}$.  With the plate scale of the WF chips, this
orbital motion is equal to $\lapprox 10$ pixels hr$^{-1}$, or $\sim1$
pixel per ten minute exposure, or up to 300 pixels in
30 hours.

The data were reduced using the following procedures: We first
produced a median sum of all 34 exposures (scaled by the mode
of the central $500\times500$ pixels).  This left a high
signal-to-noise ratio (SNR) image with all of the stars and galaxies
but with no Kuiper belt objects and no radiation events.  Then,
this median sum image was normalized to match the background of each
individual image and subtracted from each.  The difference
images were renormalized to their original background values.
At this point, we had 34 images with just
radiation events, moving objects, and noise (and no stars or
galaxies).

The next step was to combine the images so that only the Kuiper belt
objects remained.  First, we shifted the images so that a Kuiper belt
object appeared in the same pixel in all of the frames.  To accomplish
this, we specified 154 valid (see CLSD for a definition) Kuiper belt
orbits and computed the drift rate of an object
moving on that orbit during the interval of our observations.  We
shifted each image, to a fraction of a pixel, to compensate
for the drift rate to produce 34 images which were co-aligned for
the desired orbit.  Then we combined the shifted images into a set of
6 medians. Each median contained half of the images.  Since we used
the shifted images, a Kuiper belt object with the predicted drift rate
would be in the same pixel for each of these images, and thus would
remain.  In contrast, random radiation events would be
unlikely to co-align. The resulting medians were then fed to an
automatic search routine that examined each of the combined images for
faint objects.  See CLSD for a detailed description of our reduction
techniques and search algorithm.

In order to understand our limiting magnitudes, we implanted
artificial objects in the original 34 WFPC2 exposures.  The total
number of counts, $n$, that
an object of magnitude $m_{606W}$ produces in an exposure of length
$t$ through the F606W filter is (Leitherer et al$.$~1995) $$ n = t\,
10^{0.4(22.933-m_{606W})}.\eqno{(1)}$$ The conversion between
$m_{606W}$ and Johnson $V$ has been given by Holtzman et al$.$ (1995) as
$$V = m_{606W} + 0.253(V-I) + 0.012(V-I)^2. \eqno{(2)}$$ Using $V-I$
for the Sun ($+0.81$), we have $V = m_{606W} + 0.21$.  All magnitudes
quoted below are Johnson $V$ magnitudes and are based on this
calibration.

It is important to note that BKL used Equation~(1) in order to
calculate the signal of their model and that they did not include the
color terms given in Equation~(2).  Thus, their signal was too small
by $0.21$ magnitudes or by approximately 20\%.  This error corresponds
to an overestimate in the number of false detections of a factor of
approximately $2.5$.  Although this error in BKL was significant, it
cannot nearly explain the differences between our empirical results
and their model.   We shall discuss this point again below.

We implanted artificial objects of known magnitude into each of the
original 34 exposures using the IRAF routine `MKOBJECTS'.  The PSF was
assumed to be a Moffat distribution with a full width at half maximum
of 1.3 pixels and a $\beta=2.5$.  We verified that this PSF is
reasonable for our purposes by showing that the fraction of light that
falls within our photometry aperture (a five pixel plus) was indeed
the same as that from the actual HST PSF as determined by Biretta et
al$.$~(1996).  Poisson noise was added to each of the pixels in the
PSF.  Each object was implanted at a different sub-pixel location in
each exposure to simulate it moving on an arbitrary Kuiper belt orbit.

It is important to our argument that we know that MKOBJECTS performed
correctly.  Unfortunately, the objects that were implanted in the
original WFPC2 exposures were much too faint to be detected in an
individual exposure because of the large number of radiation events.
Thus, we performed the following test.  First, we fabricated an image
with the same dimensions as a WFPC2 image except that each pixel had
the same value (no noise).  We implanted artificial objects into this
image using the procedures outlined above (including noise).  We then
calculated the number of counts in each of our artificial objects
using aperture photometry.  The dots in Figure~1
show the mean number of counts for our artificial objects as a
function of $V$ magnitude. The solid curve shows the expected value as
determined by combining Equations~1 and 2.  Clearly, MKOBJECTS worked
properly.

Satisfied that our artificial objects contained the correct amount of
signal, we then processed the real WFPC2 images containing the
artificial objects using the standard reduction procedure described in
CLSD.  The black squares in Figure~2 show the probability of
detection as a function of magnitude as determined in CLSD.  To
calculate these data we implanted a total of 200 artificial objects of
known magnitude into each of the original 34 exposures; 
20 magnitudes were selected ranging from $V=26.5$ to 29.5
and 10 objects were implanted at each magnitude.  We found
that the automatic search was complete to at least $V=27$ and probably
fainter.  No artificial objects fainter than $V=29.5$ were detected.
There is an approximately linear relationship between the detection
probability and $V$ between these two extremes.  Following
Harris(1990), we defined the ``limiting magnitude'' of our search as
$V=28.6$, corresponding to a detection probability of 50\%.

In order to insure that there was no bug in our codes or our IRAF
scripts, the original CLSD team asked Peter Tamblyn to independantly
reproduce this analysis.  He used the same algorithm as CLSD, but
wrote his own codes and IRAF scripts.  He ran 165 experiments.  In
each he implanted 20 objects of constant magnitude.  His results are
presented as the gray circles in Figure~2.  He found that the search
was complete to slightly fainter than $V=27.5$.  The limiting
magnitude (50\% detection probability) was found to be $V=28.4$.
Although this is $0.2$ magnitudes brighter than found by CLSD, an
examination of Figure~2 show that this independent new analysis is in
good agreement with that of CLSD.  The difference in the two
limiting magnitudes is statistical in nature.

   The values presented in Figure~2 were calculated assuming that the
orbit of the Kuiper belt objects are precisely known.  As such, it is
an accurate representation of the limiting magnitude of the images
convolved with our reduction techniques.  It does not, however, take
into account the fact that real objects could be slightly smeared in
the images due to the finite grid spacing of our search orbits (see
CLSD for a discussion).  Thus, we performed four experiments where we
implanted 20 objects of fixed V magnitude into our HST images using a range of
orbits that were slightly offset from the one we used to search for
them.  In each experiment, we used 9 orbits which where uniformally distributed
throughout one quarter of an orbit grid.  The results of these experiments are
shown as stars in Figure~2.  The calibration shows
that for $V=27.9$ objects, $\sim50\%$ are discovered if 
smearing is included.  Even with these trailing losses included, objects
at $V=28.6$ are detected.

BKL's argument against the detection of $V=28.6$ objects is that the
SNR from such objects would be so small that the number of false
detections (predicted to be 10,000 per orbit per WF chip
\footnote{The prediction should be $\sim4000$ when corrected for BKL's
magnitude zeropoint error.})  would swamp the detection of real
objects.  In Figure~3 we plot the number of false detections that we
actually found in our final images derived from CLSD's retrograde
orbit set, as a function of magnitude.  The number is $\sim 10^{-4}$
smaller than BKL's prediction \footnote{Although BKL's prediction of
the number of false detections is several orders of magnitude larger
than we observe, it is important to note that it only represents a
difference of approximately a factor of 2 in SNR!}.

BKL attempted to explain the inconsistency between their models and
our results by arguing that our reduction procedures artificially
smooth the data.  Therefore, they contend, our procedures would
artificially decrease the noise and number of false detections.  BKL
argued that this smoothing would also affect the signal so that
$V=28.6$ objects would be undetectable.  However, the experiment we
presented in the last paragraph proves that BKL's argument cannot be
true since $V=28.6$ objects are indeed detectable!  In particular,
they are detectable in the same images that we measured a low number
of false detections!  Another (and more correct) explanation for the
difference between BKL's model and our empirical analysis is that
BKL underestimated the SNR.

\section{Discussion}

A remaining question is: `Why did BKL drastically underestimate the
SNR?'  A complete discussion of this is beyond the scope of this
paper.  However, there were several ways in
which BKL's model differed from what was actually done in CLSD. We
combined the data using medians; they used means. 
Medians are less sensitive to the deviations of the noise distribution
from Gaussian
\footnote{BKL indicate that there
would be a  1-$\sigma$ noise in a five-pixel plus of 27.9 and that this is
a factor of $\sqrt{\pi/2}$ larger than would be found if CLSD had used
means.  Our measured $\sigma$ for five-pixel plusses formed from median
data is 23.7.  This factor, alone, changes the value of $1 - f(x)$
for the Gaussian distribution function from 3.67\% to 1.74\% or the
number of false detections by a factor of 2.1.  The number of false
detections is decreased more when the neglected color term is included.}. 
They assumed that
the noise followed a Gaussian distribution.  A Kolmogorov-Smirnov test of a 
median combine of 34 images shifted assuming a retrograde orbit (one where
no objects are expected) yields a probability of 0.1\% that the distribution
of the pixel values from this median combination is Gaussian.
We constructed 6 combinations of the images, each
containing half the images, and searched for objects in each 
combination.  BKL constructed only one mean.
While our six combinations are not independent, because of the nature
of the data they still discriminate well against
false detections because a high pixel must show up in most of the sums,
not just in one or two sums.
Finally, their
estimate for the number of false detections is based on the assumption
that there were $10^6$ statistically independent measurements.
However, since each measurement was constructed from 5 pixels, and an
individual pixel was included in 5 measurements, the measurements were
not statistically independent of one another.  This effect alone
will account for a significant difference in the number of false
detections.

   In order to determine whether the differences between our results
and those of BKL were due to the difference in technique, we took
our original HST images with 20 fake objects embedded in them and
processed them according to the methods of BKL.  Like BKL, we
determined the number of false detections brighter than $m_{606W} =
28.6$ ($V=28.8$), adopting BKL's value of the total counts through
our aperture.  We find that we can reproduce BKL's results using
their techniques
\footnote{ 
Interestingly, the number of false detections derived from BKL's
model is very sensitive to how the background is defined.  BKL never
explain how they defined this, so we adopted two different approaches:
$i)$ find the median of all pixels in all 34 images and $ii)$
compute the sums of the pixels in our aperture using the BKL
clipping algorithm and average these values.  These two methods yield
backgrounds which differ by $\sim0.5$ counts/pixel/image.  The results
of our search showed that BKL's method is extremely sensitive to
this variation in the background level with the number of false
detections ranging from a few to tens of thousands.  The values we
obtained bracket the prediction of BKL.  Clearly, another problem
with BKL's model is that it is {\it extremely} sensitive to the
exact methods used.}.

Since we also reproduce our original CLSD results when we use
CLSD's methods, we conclude that the differences between our two
techniques account for the differences in the results.  Therefore, we
also conclude that BKL's conclusions are faulty because their model
did not accurately represent the HST data, nor the methods of CLSD.

\section{Concluding Remarks}

   We have shown with simple, straightforward, empirical methods that
the statistical detection of a population of small
($\sim10\,$km-sized) Kuiper belt objects by CLSD is indeed
reasonable, despite the claims of BKL to the contrary.  We find that
the limiting magnitude of the CLSD images (as defined by the 50\%
detectability limit) is $V=28.4$ 
and that objects can be detected down to at least $V = 29$. 
In addition, we find that $\sim50\%$ of the moving Kuiper belt objects
with $V=27.9$ are detected when trailing losses are included.
These results were arrived at by implanting artificial objects of known
magnitudes in the images, and re-reducing the images using the
original CLSD algorithms.  As a check, we performed this
experiment with two completely different sets of software.

We wish to emphasize several aspects of the test performed in this
paper.  First and foremost, the artificial objects were implanted in
the original 34 HST images before we started our reduction procedures
(immediately after the pipeline processing).  These were the images
that BKL used to estimate the noise in the first step of their
modeling.  As shown in Figure~1, the number of counts in our
artificial objects is consistent with BKL's estimate (correcting
for the color zero-point).  BKL claimed to have propagated these
estimates of signal and noise through our data reduction techniques
and determined that the SNR of a 28.6 magnitude object is too small
for the object to be detected.  Here, we took what is effectively
BKL's signal and noise, actually ran these numbers through our data
reduction techniques and found objects at 28.6.  In the same data that
we detected $V=28.6$ objects, we found only a small number of false
detections at the same count level, not the large number predicted by
BKL.  Note that our data reduction process is totally automated.
Thus, the results of these experiments have not been influenced by the
knowledge that there were artificial objects in the data.

As noted above, though we differ with BKL by orders of magnitude on the
number of false detections, we only need to disagree by a factor of
two on the signal/noise.  We have identified several places where this
discrepancy in SNR could arise, such as the neglect of the color term
of equation 2 by BKL (0.21 mag), the assumed noise characteristics of the data
after application of CLSD data reduction techniques,
and the lack of independence
of the pixels in the plus.  All these factors together can explain
why the model of BKL does not match the procedures of CLSD.

\acknowledgements

 We would like to thank C. Chapman, W. Cochran, R.
Gladstone, J. Parker, and P. Weissman for comments on 
the text.  We are grateful to an
anonymous referee of BKL's original {\it Icarus} paper, who first
pointed out that CLSD's calibration experiments prove that BKL's
model is flawed.

\newpage

\section{Figure Captions}

\figcaption{
The total number of counts in an object in an individual HST WFPC2
field as a function of V magnitude.  The exposure time was 500 sec.
The dots represent the mean of a set of 10 artificial objects at each
magnitude.  These data were collected using aperture photometry as
described in the text.  The curve is the expected calibration of our
data from Leitherer et al$.$~(1995, Equation~1), corrected for the
color zero-point offset of Holtzman et al$.$~(1995, Equation~2).}

\figcaption{The probability that an artificial object is detected by
our automated
reduction software as a function of V Magnitude, as defined in
Equations~1 and 2.  The objects were placed at random, sub-pixel
locations in the field.  The black squares represent the results from
CLSD's experiment of implanting artificial objects in the data.
Each square represents an experiment where 10 artificial objects of
the same magnitude were implanted (thus the resolution is $\pm
0.1$). The gray circles represent our new calibration which resulted
from 165 individual experiments, each with 20 artificial objects objects
implanted
precisely on the orbits being searched.  The results of several of these
experiments overlap one another.  In order to illustrate this the size
of the circles represent the number of experiments it represents.
Each star represents the average value of 90 similar experiments to
estimate the probability of detecting objects when trailing
losses from a complete range of orbits are included.}

\figcaption{The number of false detections in a single image directly measured
in our reduced data as a function of V Magnitude, as defined in
Equations~1 and 2.  This curve was determined by counting the total
number of false detections in CLSD's 154 retrograde orbits and
normalizing to a single image by dividing by 154.}

\epsfysize=8.3in \epsfbox{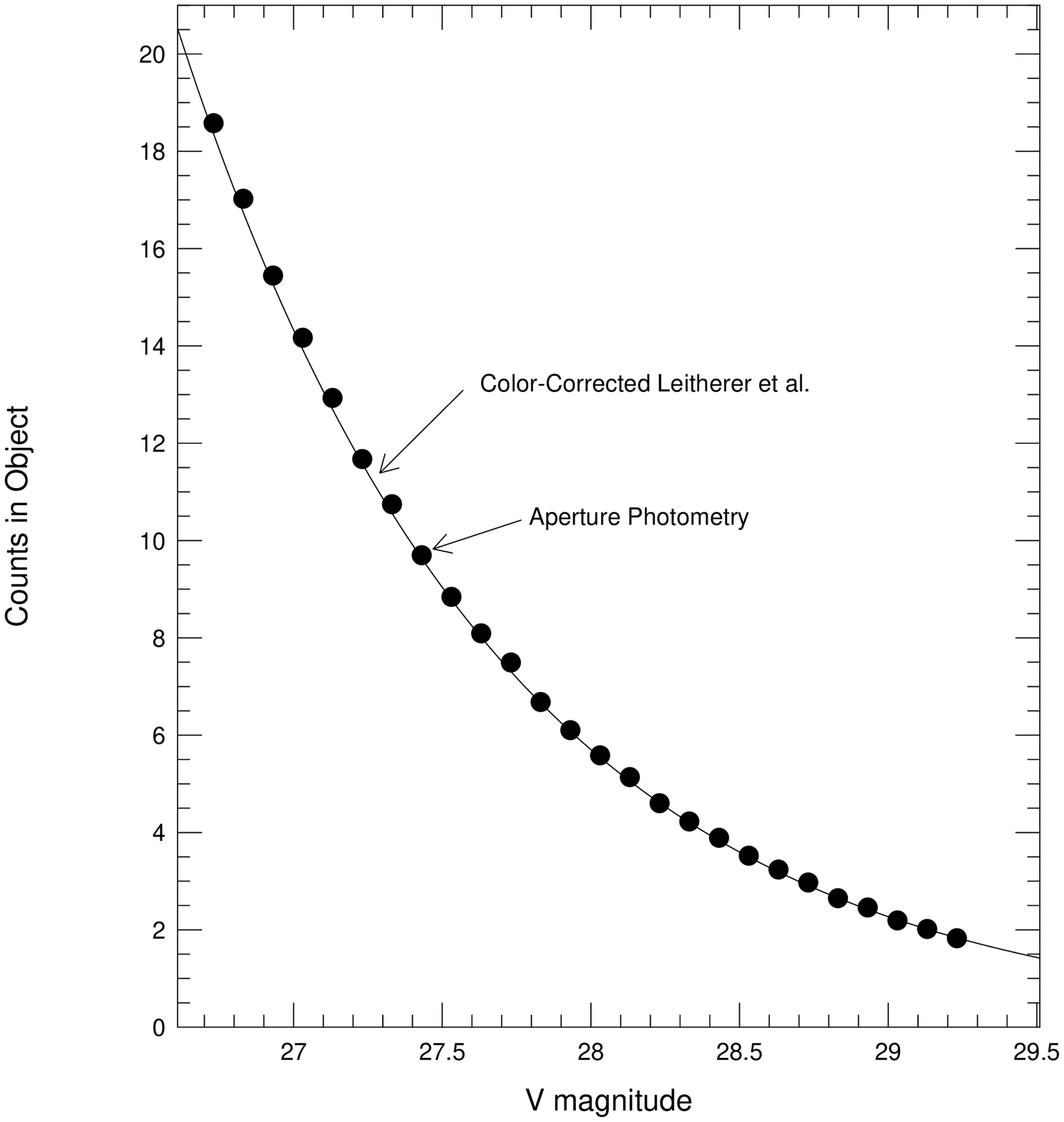}
\epsfysize=8.3in \epsfbox{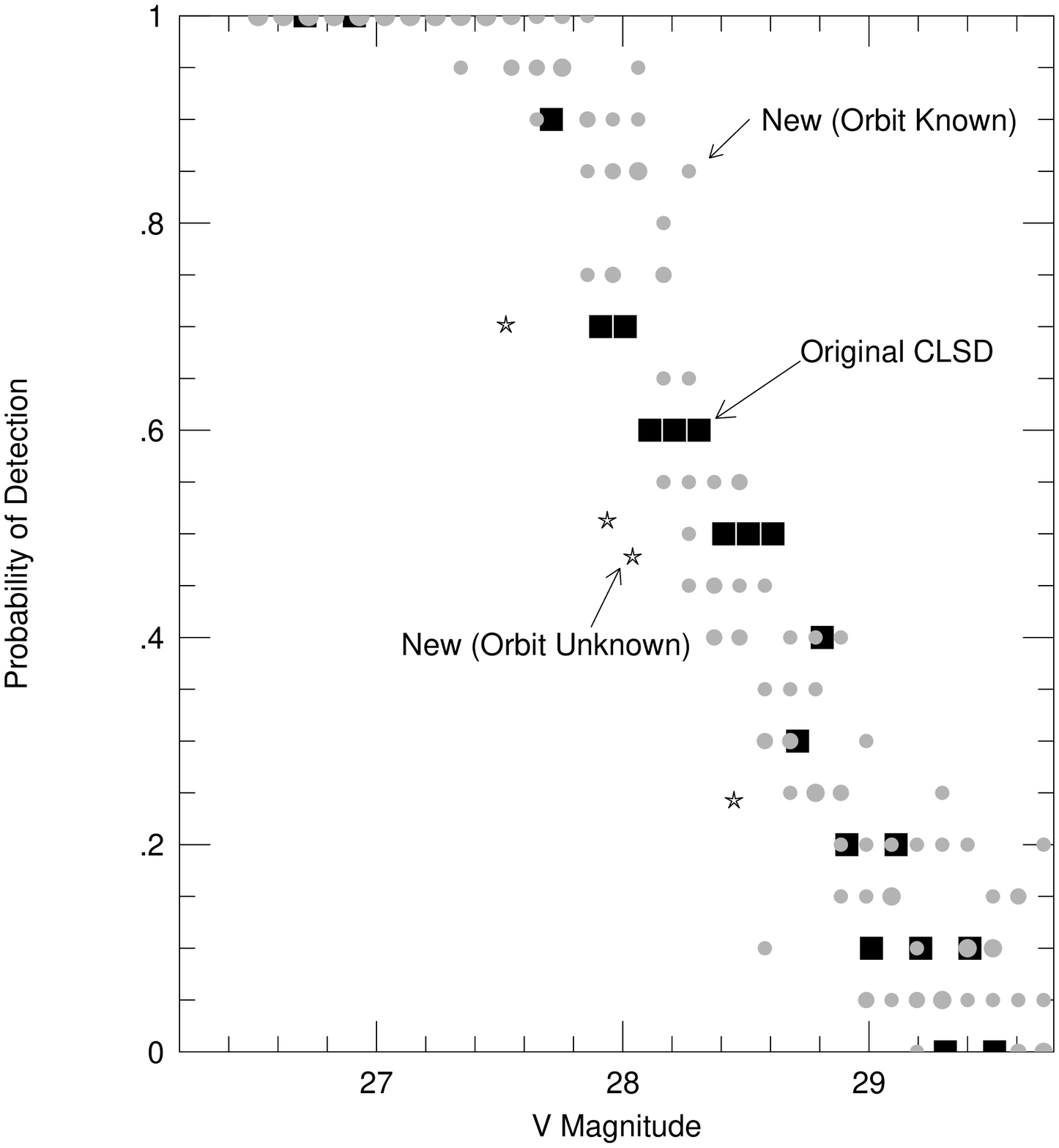}
\epsfysize=8.3in \epsfbox{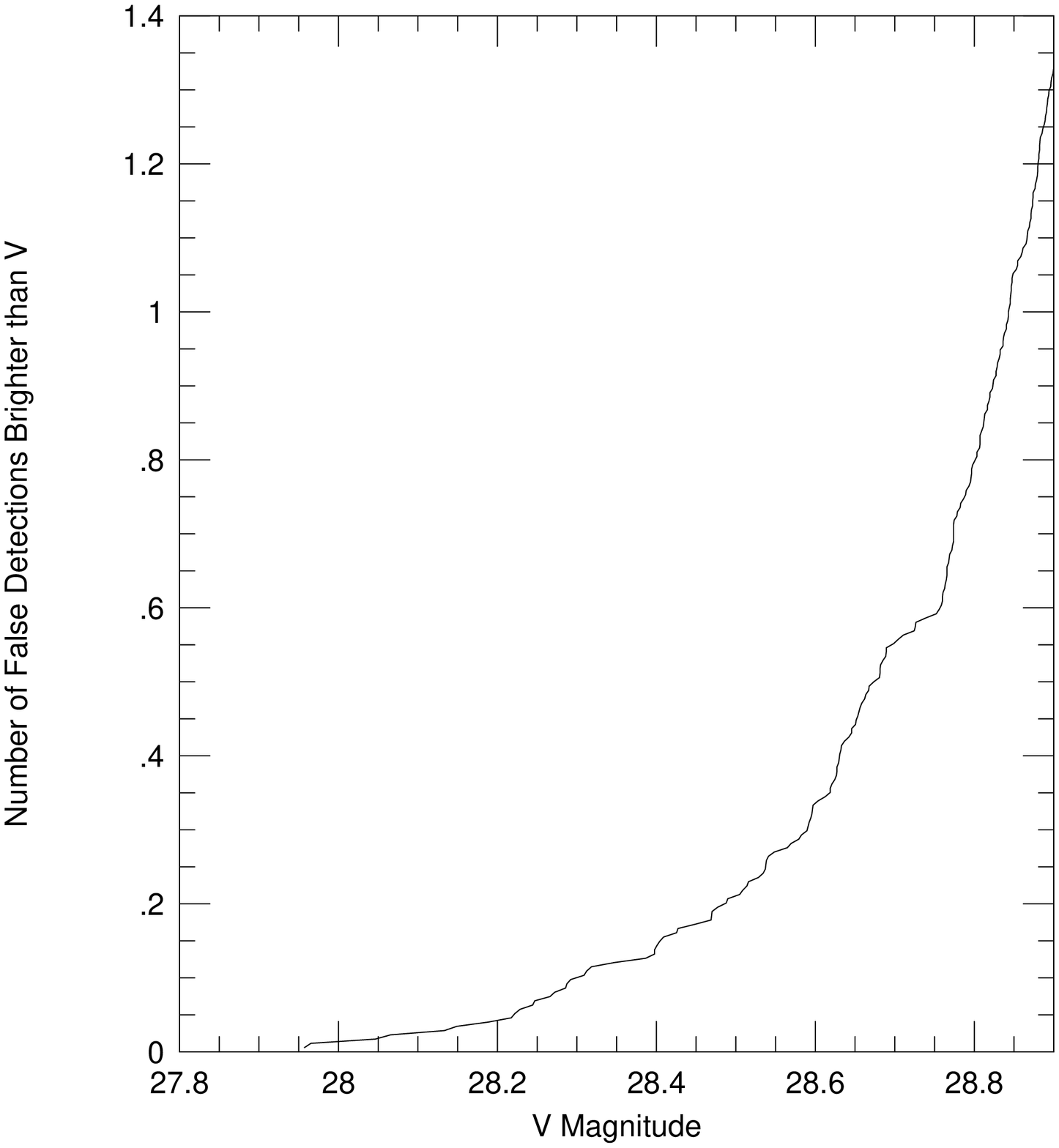}


\begin{references}

\reference {Biretta1996} Biretta, J.A. {\it et al$.$} 1996. {\it WFPC2 Instrument Handbook,
Version 4.0} (Baltimore: STScI).

\reference {Brown1997} Brown, M.E., Kulkarni, S.R., Liggett, T.J. 1997. \apjl,
{\bf 490}, L119.  (BKL)

\reference {Cochran1995} Cochran, A., Levison, H., Stern, S.A., \& Duncan, M. 1995. \apj,
{\bf 455}, 342. (CLSD)

\reference {Hainaut1994} Hainaut, O., West, R. M., Smette, A. \& Marsden, B. G.
     1994. \aap, {\bf 289}, 311

\reference {Harris1990} Harris, W. E. 1990. \pasp, {\bf 102}, 949.

\reference {Holtzman1995} Holtzman, J., {\it el al$.$} 1995. \pasp, {\bf 107}, 156.

\reference {Jewitt1996} Jewitt, D., Luu, J., \& Chen, J. 1996. \aj, {\bf 112},
1225.

\reference {Jewitt1997} Jewitt, D., Luu, J., Trujillo, C. \& Chen, J. 1997. \baas,
{\bf 29}, 1020.

\reference {Leitherer1995} Leitherer, C. 1995. {\it HST Data Handbook, Version 2.0} (Baltimore:
STScI).

\reference {Luu1988} Luu, J., \& Jewitt, D. 1988. \aj, {\bf 95}, 1256.

\reference {Mateo1988} Mateo, M. 1988. \apj, {\bf 331}, 261.

\reference {Weissman1997} Weissman, P., \& Levison, H. 1997. In {\it Pluto and Charon, }
eds. D.J. Tholen and S.A. Stern (Tucson: University of Arizona Press),
559-604.

\end{references}
\end{document}